\documentclass[12pt]{article}
\usepackage{epsfig}
\usepackage{amssymb}
\usepackage{amsmath,amsfonts,amssymb,graphicx}
\newcommand {\be}{\begin{equation}}
\newcommand {\ee}{\end {equation}}
\newcommand{\beq}{\begin{eqnarray}}
\newcommand{\eeq}{\end{eqnarray}}

\begin{document}
\vspace{-2cm}

\title{CP Violation in Neutrino Oscillations in Matter}
\author{Leonard S. Kisslinger$^{a}$, Ernest M. Henley$^{b,c}$, and
 Mikkel B. Johnson$^{d}$\\
 $^a$Department of Physics, Carnegie Mellon University, Pittsburgh, 
PA 15213 \\
 $^b$Department of Physics, University of Washington, Seattle,
WA 98195 \\
 $^c$Institute for Nuclear Theory, University of Washington, 
Seattle, WA 98195\\
 $^d$Los Alamos National Laboratory, Los Alamos, NM 87545}
\maketitle
\noindent
PACS Indices:11.30.Er,14.60.Lm,13.15.+g\vspace{0.25 in}
\begin{abstract}

We estimate CP violation for several experimental facilities 
studying neutrino oscillations. We also estimate the probability of
$\nu_{\mu}$ to $\nu_{e}$ conversion, using values of the parameter $\theta_{13}$
within the known limits,  in order to suggest new experiments to measure 
CP violation for neutrinos moving in matter.
\end{abstract}
\vspace {0.25 in}

\section{Introduction}

  The study of CP violation (CPV) is essential for understanding weak 
interactions.  Almost half a century ago CP violation in weak interactions 
was found in the decay of $K^0_L$ into $\pi^+ + \pi^{-} $ \cite{ccft64} and 
$2 \pi^0$ \cite{ckrw67}, with branching ratios of the order of .001.
The decay $K^0_L \rightarrow \pi^0 + \nu + \bar{\nu}$ is almost entirely
CP violating \cite{ll89} but requires accurate determination of the 
CKM matrix \cite{ckm63} and accurate measurements. See Ref\cite{bu08} for
a review of this experiment and references. There have many other studies 
of CP asymmetries in weak decays: see Ref\cite{blnp11} for a recent study
of $\bar{B}$ radiative decay with references to earlier work on CP violation
in various weak decays.

  In our present work we study possible CP and CPV measurements using 
neutrino oscillations. In recent years there have been a number of 
experimental studies of neutrino oscillations using neutrino beams from 
accelerators and reactors, and a most important objective of these 
experiments is the measurement of CP violation (CPV). In our present work
on estimating CPV we use parameters for the baseline and energy corresponding 
to MiniBooNE\cite{mini}, JHF-Kamioka \cite{jhf}, MINOS\cite{minos}, and
CHOOZ\cite{chooz}, which are on-going projects. 

There have been many recent studies of CP and T symmetries via neutrino 
oscillations for future facilities, e.g., see Refs\cite{dlm11,gms11}, which 
also give references to
earlier publications, and the ISS report\cite{ISS} on future neutrino
facilities. One possible future facility for studying CPV and the $\delta_{CP}$
parameter is the LBNE Project, where neutrino beams produced at Fermilab
would have a baseline of L $\simeq$ 1200 km, being detected with deep
underground detectors\cite{LBNE,LBNE07}. With the methods used in the
present work, described below, predictions of CPV with the baseline and
energies of the LBNE Project have recently been made for $\delta_{CP}$
from 90 to 0 degrees\cite{k11}. The angle $\theta_{13}$ is not well known,
and will be measured by the Daya Bay experiment in China. First, we use 
$\delta_{CP}=90^o$ and two values
of $\theta_{13}$ to explore the dependence of CP and CPV neutrino oscillations
on this parameter for all on-going projects. We then calculate CPV for
JFK-Kamioka baseline with E=.48 GeV, which has a large CPV for 
$sin\theta_{13}$=0.19 (Sec. 3), to estimate CPV for values of 
$\theta_{13}$ expected to be found (Sec. 4).

A major complication for the determination of
T, CP, and CPT violation is the interaction of neutrinos with matter as
they travel along the baseline. These matter effects have been studied by
a number of theorists. See, e.g., Refs\cite{as97,bgg97,kty02}. The main
objective of the present research is to estimate matter effects for CPV.
for the MiniBooNE, JHF-Kamioka, MINOS, and CHOOZ facilities.

For the basic interactions, which are CPT invariant for local theories, CP 
violation also implies T violation. Our present research is an extension of 
our recent work on T reversal violation\cite{hjk11}. In that study we used
the formalism of Ref\cite{ahlo01} for $\nu_e \leftrightarrow \nu_\mu$ TRV,
and that of Ref\cite{f01} for $\nu_e \rightarrow \nu_\mu$ conversion 
probability to calculate the effects of neutrinos moving through matter. 
In the present work we use the notation and formalism of 
 Jacobson and Ohlsson\cite{jo04}, who studied possible matter effects for 
CPT violation.
 
  CP violation in the $a-b$ sector is given by the transition probability,
denoted by $\mathcal{P}(\nu_a \rightarrow \nu_b)$, for a neutrino of flavor
$a$ to convert to a neutrino of flavor $b$; and similarly for antineutrinos
$\bar{\nu}_a,\bar{\nu}_b$.
   The CPV probability differences (note that the C operator changes a
particle to its antiparticle) are defined as
\beq
   \Delta\mathcal{P}^{CP}_{ab}&=& \mathcal{P}(\nu_a \rightarrow \nu_b)
-\mathcal{P}(\bar{\nu}_a \rightarrow \bar{\nu}_b) \; .
\eeq

In our present work we study $ \mathcal{P}(\nu_\mu \rightarrow \nu_e)$ and
$\mathcal{P}(\bar{\nu}_\mu \rightarrow \bar{\nu}_e)$, since the neutrino 
beams at MiniBooNE, JHF-Kamioka, and MINOS, as well as most other 
experimental facilities, are muon or anti-muon neutrinos.

\section{Transition Probability $ \mathcal{P}(\nu_\mu \rightarrow  
\nu_e)$}

In this section we review the derivation of the probability of a muon 
neutrino to convert to an electron neutrino, $ \mathcal{P}(\nu_\mu 
\rightarrow\nu_e)$, using the notation of Ref\cite{jo04}. We then make an 
estimate of the transition probalilities for
sample accelerator and reactor experiments. Although at the present time no
experiments for CPV are possible, this can serve as a basis for future
experiments. In the next section we give somewhat more accurate calculations
for CPV for the same set of experimental facilities.

As in Refs\cite{ahlo01,jo04} we use the time evolution matrix, $S(t,t_0)$ to 
derive the transition probabilities. For neutrino oscillations the initial
neutrino beam is emitted at time $t_0$, usually taken as $t_0 = 0$, and the
neutrino or converted neutrino is detected at baseline length = $L$ at
time=$t$. Since the neutrinos move with a velocity near that of the speed
of light, at the end of our derivation we take $t-t_0 \rightarrow L$, with
the units c=1.

Given the Hamiltonian, H(t), for neutrinos, the neutrino state at time = $t$
is obtained from the state at time = $t_0$ from the matrix, $S(t,t_0)$, by
\beq
             |\nu(t)> &=& S(t,t_0)|\nu(t_0)> \\
             i\frac{d}{dt}S(t,t_0) &=& H(t) S(t,t_0) \; .
\eeq

Neutrinos (and antineutrinos) are produced as $\nu_a$, where $a$ is the flavor,
$a = e,\; \mu,\; \tau$.  However, neutrinos of definite masses
are $\nu_\alpha$, with $\alpha=1,2,3$. The two forms are connected by a 3 by 3
unitary transformation matrix, $U$: $\nu_a = U \nu_\alpha$, 
where $\nu_a,\nu_\alpha$ are 3x1 column vctors and $U$ is given by 
($sin\theta_{ij} \equiv s_{ij}$)

\beq
 U=\left( \begin{array}{lcr} c_{12}c_{13} &s_{12}c_{13} & s_{13}
e^{-i \delta_{CP}} \\
     -s_{12}c_{23}-c_{12}s_{23}s_{13}e^{i\delta_{CP}} & c_{12}c_{23}-s_{12}
s_{23}s_{13}e^{i\delta_{CP}} & s_{23}c_{13} \\ 
s_{12}s_{23}-c_{12}c_{23}s_{13}e^{i\delta_{CP}} & -c_{12}s_{23}-s_{12}c_{23}
s_{13}e^{i\delta_{CP}} & c_{23}c_{13} \end{array} \right) \nonumber \; ,
\eeq
similar to the CKM matrix for quarks. We use the best fit value\cite{dlm11}
 $s_{23}=0.707$. $\theta_{13}$ is not well known. We use $s_{12} =0.56$. We use
$s_{13}=0.19$, consistent with a recent analysis\cite{gms11}, and 
$s_{13}$=0.095,
to determine the dependence of CP and CPV on this parameter, which is not 
well known. The CP phase $\delta_{CP}$ is also not well known. For 
simplicity we choose $\delta_{CP} = \pi/2$, and calculate the dependence of 
$\mathcal{P}(\nu_\mu \rightarrow\nu_e)$ on $\delta_{CP}$, as discussed below.

In the vacuum the $S(t,t_0)$ is obtained from
\beq
        S_{ab}(t,t_0)&=& \sum_{j=1}^{3} U_{aj} exp^{i E_j (t-t_0)} U^*_{bj} \; .
\eeq

  Since neutrino beams in neutrino oscillation experiments travel through
matter, and the main neutrino-matter is scattering from electrons, we must
include potential, $V = \sqrt{2} G_F n_e,$, for  neutrino electron scattering 
in the earth:
where $G_F$ is the universal weak interaction Fermi constant, and $n_e$ is 
the density of electrons in matter. Using the matter density $\rho$=3 gm/cc, 
the neutrino-matter potential is $V=1.13 \times 10^{-13}$ eV.

  The transition probability $ \mathcal{P}(\nu_\mu \rightarrow\nu_e)$
is obtained from $S_{12}$, with \\
$ \mathcal{P}(\nu_\mu \rightarrow\nu_e) = |S_{12}|^2=Re[S_{12}]^2+Im[S_{12}]^2$,
with
\beq
\label{S12}
     S_{12} &=& c_{23} \beta -is_{23} a e^{-i\delta_{CP}} A \\
     a &=&  s_{13}(\Delta -s_{12}^2 \delta) \\
     \delta &=& \delta m_{12}^2/(2 E) \\
      \Delta &=&  \delta m_{13}^2/(2 E) \\
     A & \simeq & f(t) I_\alpha* \\ 
       I_\alpha* &=& \int_0^t dt' \alpha^*(t')f(t') \\
     \alpha(t) &=& cos\omega t -i cos 2\theta sin \omega t \\
          f(t) &=& e^{-i \bar{\Delta} t} \\
 2 \omega &=& \sqrt{\delta^2 + V^2 -2 \delta V cos(2 \theta_{12})} \\
         \beta &=& -i sin2\theta sin\omega L \\
         \bar{\Delta} &=& \Delta-(V+\delta)/2 \\
            sin 2\theta&=& s_{12} c_{12} \frac{\delta}{\omega}  \; ,
\eeq
where the neutrino mass differences are $\delta m_{12}^2=7.6 
\times 10^{-5}(eV)^2$ and $\delta m_{13}^2 = 2.4\times 10^{-3} (eV)^2$. 
Note that $\delta \ll \Delta$, and $t\rightarrow L$ for $v_\nu\simeq c$.
From Eqs.(\ref{Pue},\ref{S12}):
\beq
\label{Pue}
         Re[S_{12}] &=& s_{23}a[cos((\bar{\Delta}+\delta_{CP})L)Im[I_{\alpha*}]
-sin((\bar{\Delta}+\delta_{CP})L) Re[I_{\alpha*}] \nonumber \\
         Im[S_{12}] &=& -c_{23}sin2\theta sin\omega L -s_{23} a
[cos((\bar{\Delta}+\delta_{CP})L) Re[I_{\alpha*}] \nonumber \\
        &&+sin((\bar{\Delta}+\delta_{CP})L) Im[I_{\alpha*}]] \\
    \mathcal{P}(\nu_\mu \rightarrow\nu_e) &\simeq& (c_{23}s_{12}c_{12}
\label{CPue}
(\delta/\omega) sin\omega L)^2 +(s_{23} s_{13} sin\bar{\Delta} L)^2 \\
  &&+(s_{23} s_{13} (cos\bar{\Delta} L-1.))^2 +2s_{13}s_{12}c_{12}s_{23}c_{23}
\nonumber \\
  && (\delta/\omega) sin\omega L (cos\bar{\Delta} L-1) \nonumber \;.
\eeq
From Eq(\ref{CPue}) we obtain the results for $\mathcal{P}(\nu_\mu 
\rightarrow\nu_e)$ shown in Fig.1.
\clearpage

\begin{figure}[ht]
\begin{center}
\epsfig{file=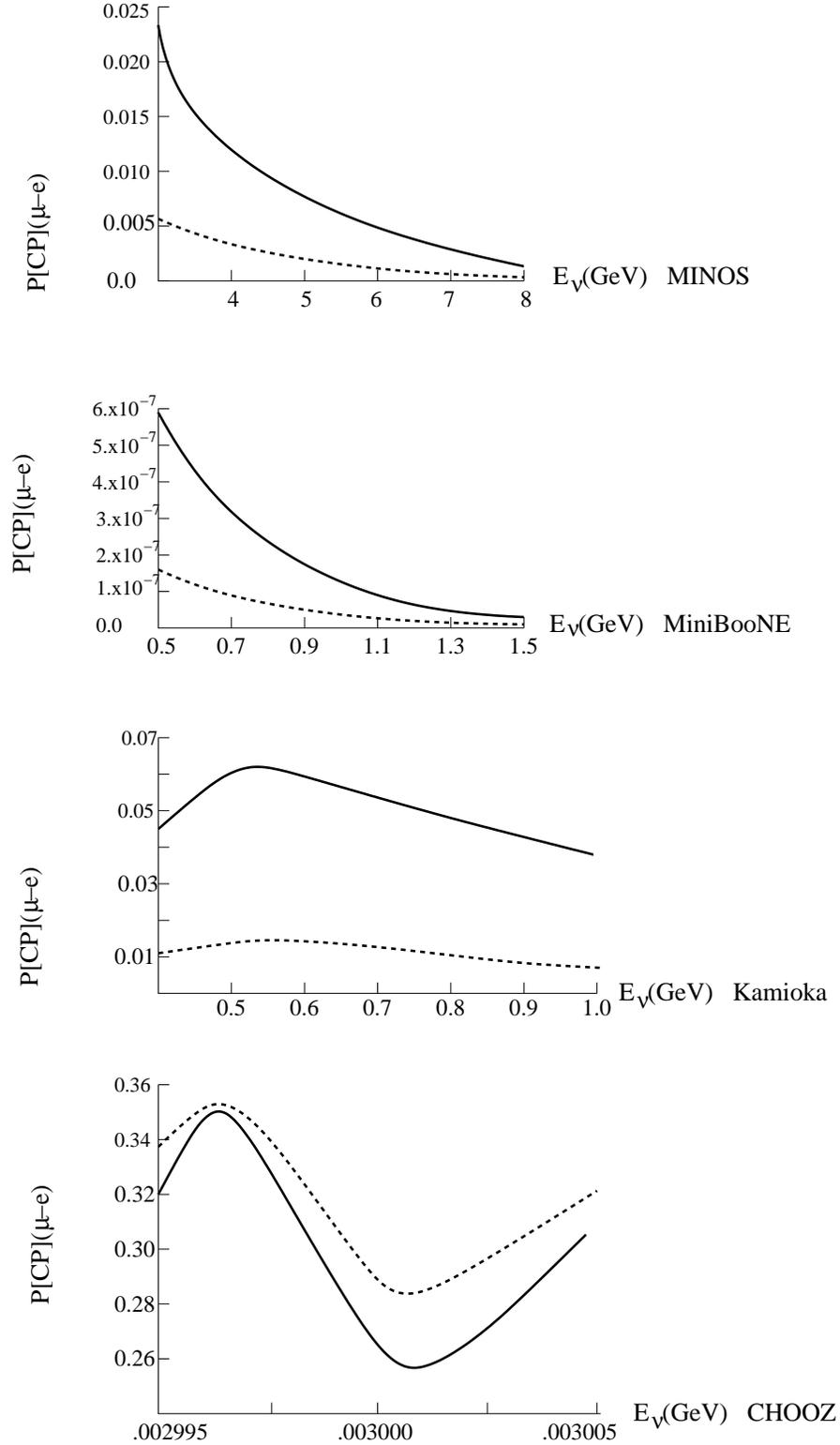,height=18cm,width=12cm}
\end{center}
\caption{\hspace{5mm} The ordinate is $\mathcal{P}(\nu_\mu \rightarrow\nu_e)$ 
for MINOS(L=735 km),
 MiniBooNE(L=500m), JHF-Kamioka(L=295 km), and 
CHOOZ(L=1 km).\hspace {5mm} 
 Energy= E in GeV. Solid curve for $s_{13}$=.19 and dashed curve for 
$s_{13}$=.095}
\end{figure}

\clearpage

 Although these results are only for muon
to electron neutrino conversion, they can provide guidance for future 
experiments on CPV via $\nu_\mu \leftrightarrow \nu_e$ oscillation. Note that
in Ref\cite{jhf} $\mathcal{P}(\nu_\mu \rightarrow\nu_e)$ was calculated for
the 295 km JHF-Kamioka project for E=0-2 GeV, and our calculation based on
the theory developed in Refs.\cite{jo04,ahlo01}, finds that with similar 
parameters $\mathcal{P}(\nu_\mu \rightarrow\nu_e)$ is in agreement for 
E=.4-1.0 GeV with this earlier estimate.
Since there is uncertainty in the value of $\delta_{CP}$, we calculated
$\mathcal{P}(\nu_\mu \rightarrow\nu_e)$ for $\delta_{CP}=0$.
We do not show the results, as they are almost the same as shown in Fig. 1.

\section{CP Violation $\Delta \mathcal{P}^{CP}_{\mu e}$}

   In this section we shall extend the derivation of the transition
probability $\mathcal{P}(\nu_\mu \rightarrow \nu_e)$ of the previous
section to derive the CPV probability
\beq
\label{CPV}
  \Delta\mathcal{P}^{CP}_{\mu e} &=& \mathcal{P}(\nu_\mu \rightarrow \nu_e)
-\mathcal{P}(\bar{\nu}_\mu \rightarrow \bar{\nu}_e) \nonumber \\
         &=&  |S_{12}|^2- |\bar{S}_{12}|^2 \,
\eeq
with $S_{12}$ defined in Eq(8) and
\beq
\label{S12bar}
  \bar{S}_{12} &=&  c_{23} \bar{\beta} -is_{23} a e^{i\delta_{CP}} \bar{A}\; ,
\eeq
with $\bar{\beta}= \beta (V \rightarrow -V)$ and $\bar{A}= A(V \rightarrow -V)$.
For example (see Eqs(16,18)) $2\bar{\omega}= \sqrt{\delta^2 + V^2 +2 \delta V 
cos(2 \theta_{12})}$ and $\bar{\bar{\Delta}}=\Delta+(V-\delta)/2$. Using
conservation of probabiltiy\cite{jo04}, $|A|^2=|\bar{A}|^2$. With
 $\delta_{CP}=\pi/2,\; e^{(-/+)i\delta_{CP}}=(-/+)i$.
\beq
\label{DCPV}
  \Delta\mathcal{P}^{CP}_{\mu e} &=& c_{23}^2(|\beta|^2-|\bar{\beta}|^2)
-2 c_{23} s_{23} a (Im[-i\beta  A^*]-Im[i \bar{\beta} \bar{A}^*]) \; .
\eeq

From Eq(\ref{DCPV}), the definitions in the previous section, defining
$s \equiv sin(\omega L)$, \\
$c \equiv cos(\omega L)$, and using $\delta_{CP} = \pi/2$
one finds
\beq
\label{DCPVf}
  \Delta\mathcal{P}^{CP}_{\mu e} &=& c_{23}^2 s_{12}^2 c_{12}^2 \delta^2
(\frac{s^2}{\omega^2}-\frac{\bar{s}^2}{\bar{\omega}^2}) +2 c_{23}s_{23}
s_{12}c_{12}s_{13}\delta (\Delta-\delta s_{12}^2)  \\
  &&(\frac{s}{\omega}(c-cos\bar{\Delta}L)\frac{\bar{\Delta}-\omega cos 2\theta}
{\bar{\Delta}^2-\omega^2}+\frac{\bar{s}}{\bar{\omega}}(\bar{c}-
cos\bar{\bar{\Delta}}L) \frac{\bar{\bar{\Delta}}-\bar{\omega} cos 2\bar{\theta}}
{\bar{\bar{\Delta}}^2-\bar{\omega}^2}) \nonumber \; .
\eeq

The results for $\Delta\mathcal{P}^{CP}_{\mu e}$ are shown in Fig.2.
Note that the largest values for CPV are for CHOOZ, with a small baseline
and low energy. With Kamioka parameters, CPV would also be a few percent,
which might be measurable. However, for experimental tests of CPV one 
needs both $\nu_\mu$ and $\bar{\nu}_\mu$ beams with the same parameters. 
Perhaps this will be possible in the future.

As has been stated in many publications, in vacuum 
$\Delta\mathcal{P}^{CP}_{\mu e}$ is given by 
$\Delta\mathcal{P}^{T}_{\mu e}$, and both vanish if $\delta_{CP}=0$.
However, with matter effects ($V \ne 0$), $\Delta\mathcal{P}^{CP}_{\mu e}$
and $\Delta\mathcal{P}^{T}_{\mu e}$, must be treated separately, as we have done. 
The magnitude of $\delta_{CP}$ is important for predictions, and it is
expected that in the future it will be determined with greater accuracy.

\clearpage
\
\vspace{6cm}

\begin{figure}[ht]
\begin{center}
\epsfig{file=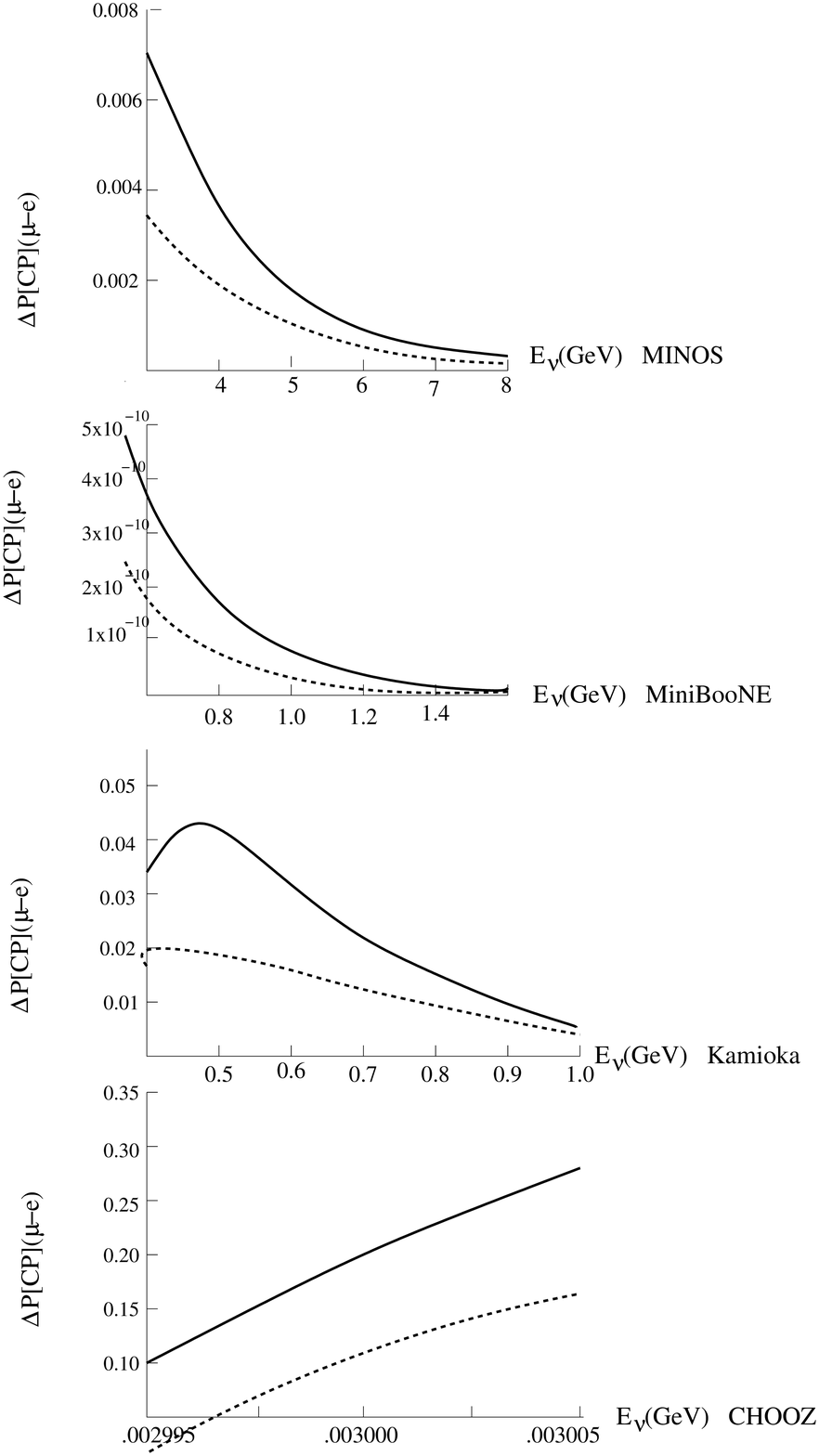,height=12cm,width=13cm}
\end{center}
\caption{The ordinate is $\Delta \mathcal{P}(\nu_\mu \rightarrow\nu_e)$ for 
MINOS(L=735 km),
 MiniBooNE(L=500m), JHF-Kamioka(L=295 km), and CHOOZ(L=1 km).\hspace{5mm}
Energy=E in GeV. Solid curve for $s_{13}$=.19 and dashed curve for 
$s_{13}$=.095}
\end{figure}

\clearpage

\section{$\Delta \mathcal{P}^{CP}_{\mu e}$ For JHF-Kamioka with
E=0.48 GeV and $sin\theta_{13}$=0.0 to 0.19}

From Fig. 2 one sees that with the value $sin\theta_{13}=0.19$ CPV reaches
a value of over 4\%, for the JFK-Kamioka project, which could be detected 
if beams of both neutrinos and antineutrinos were available. Since the
value of $sin\theta_{13}$ is not known at the present time, we use the
energy and baseline values of E=0.48 GeV, L=295 km with $sin\theta_{13}$
from 0.0 to 0.19\cite{gms11}. Using Eq.(\ref{DCPVf}), with $sin\theta_{13}$ 
a variable, one obtains the results are shown in Fig. 3.
\vspace{-2cm}

\begin{figure}[ht]
\begin{center}
\epsfig{file=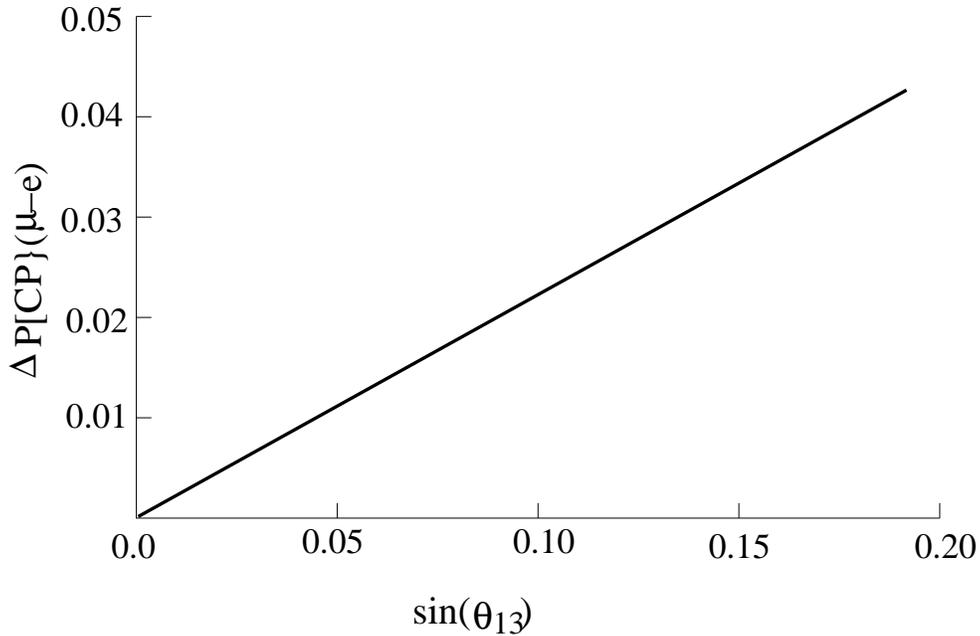,height=12cm,width=13cm}
\end{center}
\caption{$\Delta \mathcal{P}(\nu_\mu \rightarrow\nu_e)$ for 
 JHF-Kamioka(L=295 km), E=0.48 GeV, as a function of $sin\theta_{13}$}
\end{figure}

Note that since the second term in Eq.(\ref{DCPVf}) is dominant, the
dependence of $\Delta \mathcal{P}(\nu_\mu \rightarrow\nu_e)$ on 
$sin\theta_{13}$ is almost linear.

\clearpage
\section{Conclusions} 

We have estimated CP violation for a variety of experimental neutrino beam
facilities. No experiments are possible now to test CPV via neutrino 
oscillations, since beams of both neutrino and antineutrino with the same
flavor would be needed, with parameters chosen for a CPV of 1\% or more
to make the experimental measurement possible. Our results
should help in planning future experiments. Note, however, that our results
depend on the value of $s_{13}$ and $\delta_{CP}$, which are not well known,
and we have used two values for $s_{13}$ to determine the dependence of CP
and CPV on this perameter (see Fig. 2), and estimated CPV as a function 
of $s_{13}$ for the JFK-Kanioka baseline of 295 km and energy E=0.48 Gev, 
with the CPV probability over 4\% for $s_{13}$=0.19 (see Fig. 3).

\vspace{3mm}

\Large{{\bf Acknowledgements}}\\
\normalsize
This work was supported in part by the NSF grant PHY-00070888, in part 
by the DOE contracts W-7405-ENG-36 and DE-FG02-97ER41014.

\end{document}